\documentclass[12pt]{article} 
\usepackage{psfrag,graphicx}
\usepackage{multicol}
\usepackage{amsmath}
\usepackage{amsfonts}
\usepackage{amssymb}
\usepackage{enumitem}
\usepackage{color,xcolor,boxedminipage}
\usepackage[misc,geometry]{ifsym}
\usepackage{epstopdf}
\usepackage{float}

\begin{document}

\hspace{12pt}

\huge
 \begin{center}
Robust Principal Component Analysis \\ for Compositional Tables\\ 

\hspace{24pt}\bigskip

\large
J. Rendlov\'a$^1$, K. Hron$^1$, K. Fa\v cevicov\'a$^1$ and P. Filzmoser$^2$ \\

\hspace{18pt}\bigskip

\small  
$^1$ Department of Mathematical Analysis and Applications of Mathematics, \\
Faculty of Science, Palack\'y University, Olomouc, Czech Republic;\\

\smallskip

$^2$ Institute of Statistics and Mathematical Methods in Economics, \\ Vienna
University of Technology, Vienna, Austria.


\end{center}

\hspace{12pt}

\normalsize

\section*{Abstract}

A data table which is arranged according to two factors can often be considered as a~compositional
table. An example is the number of unemployed people, split according to gender and age classes. Analyzed as compositions, the relevant information would consist of ratios between different cells of such a table. This is particularly useful 
when analyzing several compositional tables jointly, where the
absolute numbers are in very different ranges, e.g.~if unemployment data are considered from different countries. Within the framework of the logratio methodology, compositional tables can be decomposed into independent and interactive parts, and orthonormal coordinates can be assigned to these parts. However, these coordinates usually require  
some prior knowledge about the data, and they are not easy to handle for exploring the relationships between the given factors. 

Here we propose a special choice of coordinates with a direct relation
to centered logratio (clr) coefficients, which are particularly useful for 
an interpretation in terms of the original cells of the tables. With these
coordinates, robust principal component analysis (PCA) is performed for dimension reduction, allowing to investigate the relationships between the factors.
The link between orthonormal coordinates and clr coefficients enables to apply 
robust PCA, which 
would otherwise suffer from the singularity of clr coefficients.

\hspace{12pt}

\noindent\textbf{Keywords:} Compositional data; Compositional table; Robust principal component analysis; Independence table; Interaction table; Pivot coordinates.

\clearpage

\section{Introduction}
\label{intro}
Many practical data sets (e.g. in economics \cite{facevicova14,facevicova16}; 
biology \cite{herder08,dickhaus12}; 
or sociology \cite{egozcue08,ortego16}) 
consist of observations which contain inherently relative information about the distribution according to two factors.
From a mathematical perspective, this leads to a two-factorial extension of vector compositional data \cite{aitchison86,pawlowsky15} 
carrying information about a relationship between and within these (row and column) factors.    
In \cite{egozcue08}, it was shown that such a structure, called a compositional table $\boldsymbol{x}$,
\begin{equation} 
\label{codatable}
\boldsymbol{x} = 
	\begin{pmatrix} 
		x_{11} & \cdots & x_{1J} \\ 
		\vdots & \ddots & \vdots \\ 
		x_{I1} & \cdots & x_{IJ}  
	\end{pmatrix},
 x_{ij}>0, i = 1, ..., I, j = 1, ..., J,
\end{equation}

\noindent can be decomposed into its independent and interactive parts where their separate analysis can be advantageous for further interpretation concerning both factors and their relationships. Similar as compositional data, compositional tables are driven by the Aitchison geometry \cite{pawlowsky01} 
which inhibits a direct
application of multivariate statistical tools. 
Instead, an appropriate coordinate representation needs to be established first to map the information 
from the Aitchison geometry into the real space. 
Such a coordinate system and a~decomposition into independent and interactive parts were proposed in \cite{facevicova16}.

A common primary task in multivariate statistics is to reduce the dimensionality of the data at hand, done using principal component analysis (PCA). As stated above, in case of compositional data, and consequently also compositional tables, this needs to be done in a proper coordinate representation that maps the Aitchison geometry of compositions to a standard Euclidean geometry \cite{pawlowsky01}. 
To eliminate the influence of outlying observations in PCA, 
in \cite{filzmoser09} it was proposed to estimate the covariance matrix for robust PCA by the Minimum Covariance Determinant
(MCD) estimator \cite{maronna06}. 
Since centered logratio (clr) coefficients \cite{aitchison86}, 
that aggregate relative information on single compositional parts, lead to singularity and are not appropriate for most robust methods including the MCD estimator, loadings and scores of PCA need to be computed from isometric logratio (ilr) coordinates \cite{egozcue03} 
of the compositional data and then transformed back to clr coefficients for a better interpretation of the resulting compositional biplot. 

Accordingly, the aim of this paper is to generalize the previous considerations on dimension reduction of vector compositional data and to propose a robust approach to principal component analysis of compositional tables. In Sections~\ref{sec:2} and~\ref{sec:3}, the logratio methodology of vector compositional data and their dimension reduction using principal component analysis, respectively, are briefly reviewed. Compositional tables and their interpretable processing using robust PCA are introduced in Section~\ref{sec:4}. The new methodology is illustrated in Sections~\ref{sec:5} and~\ref{sec:6} on real data sets from OECD Statistics using the statistical software R, namely the robCompositions package. 
Data from several different countries containing unemployment information with gender distribution and age structure are processed as a set of $2 \times 4$ compositional tables. Therefore, a robust compositional biplot is a possible tool to analyze the distribution of unemployment rates in these countries as well as gender and age differences. Data from the area of education, carrying relative information about fields of study and the resulting degree in given countries, are approached as larger $3 \times 8$ compositional tables, and results for men and women are compared.

\section{Logratio methodology of compositional data}
\label{sec:2}

In many applications, the relative structure of the observations is more interesting
than the absolute values of their components. For example, when considering numbers of students attending bachelor, master and doctoral studies at different universities, the ratios among these three groups might be more relevant for a statistical analysis than just the empirical values, which might not be comparable because of different total student numbers. In other words, the actual total number of students (sufficiently high so that the impact of a measurement error with small sample sizes can be neglected) might be considered as not informative for the purpose of the analysis. However, to work with quantitatively described contributions on a given whole in a concise and meaningful manner, some concepts need to be introduced first. 

A positive (row) vector $\boldsymbol{x}=(x_1, x_2, ..., x_D)$ is defined to be a \textit{D}-part composition if it carries relative information, i.e. the ratios between the components are informative \cite{aitchison86,pawlowsky15}. 
Any compositional vectors with equal number of parts are considered to be representatives of the same \textit{equivalence class} if one vector is obtained from another by a positive scalar multiplication \cite{pawlowsky15}. 
Accordingly, equivalence classes of compositional data are represented without loss of information in a \textit{D}-part simplex,
\begin{equation} 
\mathcal{S}^D=\left\{\boldsymbol{x}=(x_1, ..., x_D) | x_i>0, i = 1, ..., D, \sum_{i=1}^D x_i = \kappa \right\} 
\end{equation} 
for any $\kappa>0$. The choice of $\kappa$ (being 1 for proportions and 100 for percentages) is irrelevant for the analysis and can also vary throughout the compositional data set. Formally, the closure operation 
\begin{equation} 
\label{closure}
\mathcal{C}(\boldsymbol{x})=\left(\frac{\kappa\cdot x_1}{\sum_{i=1}^D x_i}, \frac{\kappa\cdot x_2}{\sum_{i=1}^D x_i}, ..., \frac{\kappa\cdot x_n}{\sum_{i=1}^D x_i}\right) 
\end{equation}
can be applied to rescale the data to a given constant sum ($\kappa$) representation. Accordingly, the $D$-part simplex is a sample space of (representatives of equivalence classes of) compositions. The constant sum representation is useful, e.g. for a comparison of several compositions from a sample. As an interesting consequence, possibly deviating (outlying) observations from the main data cloud are characterized by aberrant ratios rather than by significantly high or low absolute values of components \cite{filzmoser13}. 

As mentioned above, relevant information in data with compositional nature is contained only in ratios. Consequently, results of statistical processing should not depend on the sum $\kappa$ of compositional parts and instead of Euclidean distances, relative differences are used to express distances between observations. This principle called \textit{scale invariance} is the first of three basic compositional principles \cite{pawlowsky15}. 
Often the original data contain some non-informative part(s) in the compositional vector that are not of interest. Hence, we do not expect any change of results concerning the respective subcomposition when these parts are removed from the data. 
\textit{Subcompositional coherence} is a principle stating that results obtained from a \textit{d}-part subcomposition, $d < D$, are not in contradiction with results obtained by an analysis of the original \textit{D}-part composition. Finally, \textit{permutation invariance} states that the results are independent 
from a chosen order of parts within the composition, an expectable assumption for any reasonable statistical processing.

Due to the above principles and the relative scale of compositional data, the Euclidean geometry needs to be replaced with the Aitchison geometry \cite{egozcue03}. 
Operations of \textit{perturbation} and \textit{power transformation}, being analogous to a sum of two vectors and a multiplication of a vector by a scalar in the real Euclidean geometry, are defined as
\begin{equation} 
\boldsymbol{x}\oplus\boldsymbol{y}=(x_1y_1, ..., x_Dy_D),\quad \alpha\odot\boldsymbol{x}=(x_1^\alpha, ..., x_D^\alpha), 
\end{equation}
where $\boldsymbol{x}$ and $\boldsymbol{y}$ are \textit{D}-part compositions, and $\alpha$ is a real constant. Accordingly, operations of perturbation and power transformation 
form a $(D-1)-$di\-men\-si\-on\-al vector space $(\mathcal{S}^D,\oplus,\odot)$ \cite{pawlowsky15}. 

To obtain Euclidean vector space structure, the \textit{Aitchison inner product, norm} and \textit{distance} are defined for \textit{D}-part compositions $\boldsymbol{x}$ and $\boldsymbol{y}$ as
\[
\left\langle \boldsymbol{x}, \boldsymbol{y} \right\rangle_A = \frac{1}{2D}\sum_{i=1}^D\sum_{j=1}^D \ln \frac{x_i}{x_j} \ln \frac{y_i}{y_j}, \quad
\left\|\boldsymbol{x}\right\|_A = \sqrt{\left\langle \boldsymbol{x}, \boldsymbol{x} \right\rangle_A },\] 
\begin{equation}
d_A(\boldsymbol{x},\boldsymbol{y}) = \left\|\boldsymbol{x}\ominus\boldsymbol{y}\right\|_A, 
\end{equation}
respectively, where $\boldsymbol{x}\ominus\boldsymbol{y} = \boldsymbol{x}\oplus\left[(-1)\odot\boldsymbol{y}\right]$.
 
Given the introduced specifics of compositional data endowed with the Aitchison geometry, standard multivariate statistical methods cannot be applied directly on raw data. 
It is advisable to employ the \textit{working on coordinates} principle; that is 
to represent the compositional data in real coordinates before starting with a statistical analysis \cite{pawlowsky03,mateu11}. 
There are two types of isometric coordinate representations with respect to the Aitchison geometry.
\textit{Centered logratio coefficients} \cite{aitchison86} 
are defined as
\begin{equation} 
\mathrm{clr}(\boldsymbol{x}) = \left(\ln\frac{x_1}{g(\boldsymbol{x})}, 
\ln\frac{x_2}{g(\boldsymbol{x})}, ..., \ln\frac{x_D}{g(\boldsymbol{x})}\right), 
\end{equation}
where $g(\boldsymbol{x})$ stands for the geometrical mean of the whole composition. Each clr coefficient aggregates all pairwise logratios with a given compositional part, thus enabling for a simple and meaningful interpretation in terms of dominance of that part with respect to the other parts \textit{on average}. Consequently, clr coefficients are useful for a graphical interpretation of compositional data including compositional biplots as a result of a dimension reduction using PCA \cite{aitchison02}. 
On the other hand, clr coefficients sum up to zero which leads to a singular covariance matrix, being inappropriate for processing using common robust statistical methods \cite{filzmoser09,filzmoser13}. 

Fortunately, \textit{isometric logratio coordinates} $\boldsymbol{z}\in\boldsymbol{R}^{D-1}$ representing orthonormal coordinates with respect to the Aitchison geometry can be derived as
\begin{equation} 
\boldsymbol{z} = \mathrm{ilr}(\boldsymbol{x}) = \left( \left\langle \boldsymbol{x},\boldsymbol{e}^1 \right\rangle_A, \left\langle \boldsymbol{x},\boldsymbol{e}^2 \right\rangle_A, ..., \left\langle \boldsymbol{x},\boldsymbol{e}^{D-1} \right\rangle_A \right), 
\end{equation}
where $D$-part compositions $\boldsymbol{e}^i = \mathcal{C}(e_1^i, e_2^i, ..., e_D^i)$, $i = 1, ..., D-1$, form an orthonormal basis on the simplex. 

Obviously, the interpretation of ilr coordinates might be more tricky than in the case of clr coefficients as there are infinitely many possibilities of constructing ilr coordinates depending on the choice of basis vectors $\boldsymbol{e}^i,\ i=1,\dots,D-1$. \textit{Sequential binary partitioning (SBP)} of compositional parts is one possibility for providing a meaningful choice of $\boldsymbol{e}^i$ for the practitioner which is corresponding to the prior knowledge about compositions and resulting in coordinates called balances \cite{egozcue05}. 
Moreover, there is a linear transformation between ilr coordinates and clr coefficients, done through a $D\times (D-1)$ matrix $\boldsymbol{V}$ of clr representations of the ilr basis vectors,
\begin{equation} 
\label{ilrclr}
\mathrm{clr}(\boldsymbol{x}) = \boldsymbol{Vz} = 
[\mathrm{clr}(\boldsymbol{e}_1)^T,\mathrm{clr}(\boldsymbol{e}_2)^T, ..., \mathrm{clr}(\boldsymbol{e}_{D-1})^T]\cdot \mathrm{ilr}(\boldsymbol{x}).
\end{equation} 

Recently, \textit{pivot coordinates} \cite{fiserova11,hron17} 
\begin{equation} 
z_i^{(l)} = \sqrt{\frac{D-i}{D-i+1}} \ln \frac{x_i^{(l)}}{\sqrt[D-i]{\prod_{j=i+1}^D{x_j^{(l)}}}},
\end{equation}
\noindent 
were introduced as a special case of ilr coordinates. 
They are appropriate especially in situations where no prior knowledge about 
how to perform SBP is available, e.g. in~\cite{bruno15,buccianti14,dumuid18,kalivodova15}.  
Here, $x_i^{(l)}$ refers to the \textit{i}-th part of the re-ordered composition
$(x_l,x_1, \ldots,$ $x_{l-1},x_{l+1},\ldots , x_D)$ which can be rewritten as $(x_1^{(l)},x_2^{(l)},\ldots ,
x_l^{(l)},x_{l+1}^{(l)},\ldots, x_D^{(l)})$.
In each of the $D$ coordinate systems, a permutation of compositional parts needs to be performed, so that the \textit{l}-th part of $\boldsymbol{x}$ ($l=1,\dots,D$) stands at the first position. Accordingly, the first pivot coordinate in each system, $z_1^{(l)}$, then clearly explains all relative information about part $x_l$ and, additionally, it is proportional to the respective clr coefficient
\begin{equation} 
\label{clrscaled}
\mathrm{clr}(x_l^{(l)}) = \sqrt{\frac{D}{D-1}} z_1^{(l)}.
\end {equation} 
The expression (\ref{ilrclr}) holds also for pivot coordinates which is particularly useful in the context of this paper.
 
\section{Robust principal component analysis for compositional data}
\label{sec:3}

When dealing with large-scale data sets, dimension reduction is often of primary interest. PCA is one of the widely used methods for this purpose also in a compositional approach, converting possibly correlated original variables from the data at hand into a smaller set of linearly uncorrelated variables called principal components (PCs). Additionally, the first component accounts for the largest variance of the given data, the second one for a~maximum of the remaining variance, etc., under the constraint of being orthogonal to all the previous principal components \cite{johnson07}. 

The covariance matrix $\boldsymbol{C}$ estimated from a real data matrix $\boldsymbol{X}$ can be spectrally decomposed into
\begin{equation} 
\boldsymbol{C}=\boldsymbol{GLG}^T,
\end{equation}
\noindent where $\boldsymbol{G}$ is a matrix of eigenvectors and $\boldsymbol{L}$ represents a diagonal matrix of eigenvalues of $\boldsymbol{C}$. It is then possible to define the PCA transformation as 
\begin{equation} 
\boldsymbol{X}^\ast = (\boldsymbol{X} - \boldsymbol{1t}^T) \boldsymbol{G},
\end{equation}
where $\boldsymbol{t}$ is the location estimator and $\boldsymbol{1}$ is a vector of ones with length $n$. The columns of the matrix 
$\boldsymbol{X}^\ast$, the coordinates of the principal components, are called \textit{scores} and the columns of $\boldsymbol{G}$, containing the respective basis vectors, are called \textit{loadings}. Typically, only the first few principal components are considered for further analysis. Taking into account only two PCs, a graphical outcome called \textit{biplot} can depict both loadings as arrows and scores as points in one plot, where associations can be revealed.

It is common to take $\boldsymbol{t}$ as the arithmetic mean and 
$\boldsymbol{C}$ as the sample covariance matrix, however, both are
very sensitive to outlying observations.
Robust alternatives can be obtained by using the \textit{Minimum Covariance Determinant (MCD)} estimators of location and covariance \cite{maronna06}. 
However, this approach inquires working in ilr coordinates to obtain full rank data in order to get the MCD estimate of the covariance matrix and the respective matrix of eigenvectors $\boldsymbol{G}$. 
In addition, ilr coordinates ensure subcompositional coherence and enable to keep affine equivariance of the results to the change of basis.

Accordingly, robust principal component analysis of compositional data based on the MCD estimator requires ilr coordinates $\boldsymbol{z}_i$ as an input, and the scores $\boldsymbol{z}_i^\ast$
are given by
\begin{equation} 
\boldsymbol{z}_i^\ast = \boldsymbol{G}^T(\boldsymbol{z}_i - \boldsymbol{t}).
\end{equation}
\noindent Once PCA is performed, the loadings can be transformed back to clr coefficients as
\begin{center}
\smallskip
$\boldsymbol{G}_{\mathrm{clr}} = \boldsymbol{VG}$,
\smallskip
\end{center}
accounting for compositional biplot construction with meaningful interpretation, whereas the scores remain identical and only a column of zeros is added to the end. Clr coefficients are also worth as such for their simple construction as an amalgamation of pairwise logratios of a given part. Due to the zero-sum constraint of clr coefficients, their covariance structure is distorted, thus the interpretation of the biplot in terms of the correlation between coefficients (through angles between arrows) might be misleading. Instead, the focus is on links between vertices of arrows as they stand for a proportionality between the original compositional parts \cite{aitchison02}. 
On the other hand, due to the link with pivot coordinates, the single clr variables (or the respective loadings) can be used to identify observations with a high dominance of the respective  parts in a compositional vector \cite{kynclova16}. 

\section{Compositional tables}
\label{sec:4}

Compositional tables describe quantitatively relative contributions to a given whole that is distributed according to two (row and column) factors. Mathematically, this leads to a two-factorial extension of vector compositional data, carrying information about a relationship between these factors. A compositional table (\ref{codatable}) thus can be represented, e.g. either as a contingency table (with sufficiently high numbers of counts in the cells) or as a table of the same order with ML estimates of the respective probabilities -- due to scale invariance, the relative information (contained in the ratios between the cells) is the same in both cases \cite{egozcue08,egozcue15}. 
Hence, the concept of compositional tables (\ref{codatable}) covers both the discrete case of contingency tables and its continuous counterpart (e.g. input-output tables, see \cite{facevicova14}). 
Nevertheless, in the compositional context, a concrete table represents usually just one realization in a sample from a multivariate continuous distribution.
Due to the decision to treat the data from such a distribution compositionally, the possible order of the factor categories (e.g. age levels) is ignored, making this a relevant subject for future research.

\indent Since compositional tables form a direct extension of vector compositional data, all the principles and operations introduced in Section~\ref{sec:2} apply, up to some minor modifications due to the two-factorial structure of the tables. 

Accordingly, 
the closure operation 
\begin{equation} \mathcal{C}(\boldsymbol{x}) = 
	\begin{pmatrix} 
		\frac{\kappa x_{11}}{\sum_{i,j}x_{ij}} & \cdots & \frac{\kappa x_{1J}}{\sum_{i,j}x_{ij}} \\ 
		\vdots & \ddots & \vdots \\ 
		\frac{\kappa x_{I1}}{\sum_{i,j}x_{ij}} & \cdots & \frac{\kappa x_{IJ}}{\sum_{i,j}x_{ij}}  
	\end{pmatrix}
\end{equation}
is used to represent a compositional table $\boldsymbol{x}$ in an $IJ$-part simplex $\mathcal{S}^{IJ}$ of vectorized tables 
$\mathrm{vec}(\boldsymbol{x})=(x_{11},\ldots,x_{I1},\ldots,x_{IJ})$. Perturbation $\boldsymbol{x} \oplus \boldsymbol{y}$, powering $\alpha \odot \boldsymbol{x}$, and the Aitchison inner product $\left\langle \boldsymbol{x}, \boldsymbol{y} \right\rangle_A$ of two tables $\boldsymbol{x},\,\boldsymbol{y}$ and a real number $\alpha$ can be defined analogously \cite{egozcue08,egozcue15}: 

\[ \boldsymbol{x} \oplus \boldsymbol{y} = 
\begin{pmatrix} 
		x_{11}y_{11} & \cdots & x_{1J}y_{1J} \\ 
		\vdots & \ddots & \vdots \\ 
		x_{I1}y_{I1} & \cdots & x_{IJ}y_{IJ}  
	\end{pmatrix}, \quad
 \alpha \odot \boldsymbol{x} = 
\begin{pmatrix} 
		x_{11}^\alpha & \cdots & x_{1J}^\alpha \\ 
		\vdots & \ddots & \vdots \\ 
		x_{I1}^\alpha & \cdots & x_{IJ}^\alpha  
	\end{pmatrix},
\]
\begin{equation} \left\langle \boldsymbol{x}, \boldsymbol{y} \right\rangle_A = \frac{1}{2IJ}\sum_{i,j}\sum_{k,l} \ln \frac{x_{ij}}{x_{kl}} \ln \frac{y_{ij}}{y_{kl}}. 
\end{equation}

\noindent It is straightforward to derive that the dimension of the simplex $\mathcal{S}^{IJ}$ is $IJ-1$, corresponding to the dimensionality of $(I\times J)$-compositional tables.

Permutation invariance and subcompositional coherence are valid with respect to the two factors of the compositional tables, allowing to permute and discard entire rows or columns only.

To analyze compositional tables, it is beneficial to work also with the so-called \textit{independence} and \textit{interaction} tables. These can be obtained through an orthogonal decomposition \cite{egozcue08}

\begin{equation}
\label{relindint}
\boldsymbol{x}=\boldsymbol{x}_{ind} \oplus \boldsymbol{x}_{int}.
\end{equation}

\noindent 
Here, the independence table is constructed to extract all the relative information about row and column factors under the assumption that the original compositional table is a~product of its row and column geometric marginals, and the interaction table contains information about the relationships between the row and column factors. Therefore, in case of actual independence in the data at hand (in the above sense), all the entries of the interaction table are the same, since there is no remaining information left in the data after the decomposition; the interaction table thus forms a neutral element with respect to the Aitchison geometry of compositional tables. Otherwise, the interactive part describes the nature of the deviation from an independent situation. Needless to say, analyzing each of these new tables separately allows for a deeper insight into the original data.

It turns out that the introduced decomposition can be easily derived from row and column projections of the compositional table onto marginal subspaces (for further details, see \cite{egozcue08}), 

\begin{equation} \mathrm{row}^\bot (\boldsymbol{x}) = 
\begin{pmatrix} 
		g(x_{11}, ..., x_{1J}) & \cdots & g(x_{11}, ..., x_{1J}) \\ 
		\cdots & \cdots & \cdots \\ 
		g(x_{I1}, ..., x_{IJ}) & \cdots & g(x_{I1}, ..., x_{IJ})  
	\end{pmatrix},
\end{equation}

\begin{equation} \mathrm{col}^\bot (\boldsymbol{x}) = 
\begin{pmatrix} 
		g(x_{11}, ..., x_{I1}) & \vdots & g(x_{1J}, ..., x_{IJ}) \\ 
		\vdots & \vdots & \vdots \\ 
		g(x_{11}, ..., x_{I1}) & \vdots & g(x_{1J}, ..., x_{IJ})  
	\end{pmatrix},
\end{equation} 
where $g(.)$ denotes the geometric mean of the cells in the argument and $\bot$ stands for orthogonality of the projections. 

Recalling the case of independence in probability tables, it is instant to get the independence table simply by perturbing both these projections, 
$\boldsymbol{x}_{ind}= \mathrm{row}^\bot (\boldsymbol{x}) \oplus \mathrm{col}^\bot (\boldsymbol{x})$. From (\ref{relindint}) it follows that the interaction table is just a decomposition remainder in $\boldsymbol{x}_{int} = \boldsymbol{x} \ominus \boldsymbol{x}_{ind}$. 
For practical calculations, the following formulas are used to obtain the single
entries of these tables,
\[
x^{ind}_{ij} = \left( \prod_{k=1}^I \prod_{l=1}^J x_{kj}x_{il} \right)^{\frac{1}{IJ}}\propto
\left( \prod_{k=1}^I x_{kj} \right)^{\frac{1}{I}}
\left( \prod_{l=1}^J x_{il} \right)^{\frac{1}{J}},
\]
\begin{equation}
x^{int}_{ij} = \left( \prod_{k=1}^I \prod_{l=1}^J \frac{x_{ij}}{x_{kj}x_{il}} \right)^{\frac{1}{IJ}}.
\end{equation}
It is crucial to realize that the dimensions of $\boldsymbol{x}_{ind}$ and $\boldsymbol{x}_{int}$ lower to $I+J-2$ for the independence tables, which follows immediately from the dimensions of the row and column projections being, respectively, $I-1$ and $J-1$, and to $(I - 1)(J - 1)$ for the interaction tables, which is easily obtained from the orthogonality of the decomposition.

As stated in the previous sections, a coordinate representation which respects the sample space dimensionality is needed to perform robust PCA of compositional data. In case of compositional tables (and particularly their decomposed parts), a generalization of balance coordinates needs to consider two SBPs according to each factor \cite{facevicova18}. 
Even for moderate numbers of rows and columns, the interpretation of this coordinate representation gets rather complex without a deeper expert knowledge. Therefore, only a~two-factorial alternative to pivot coordinates is usually used in practice \cite{facevicova16,facevicova18}. 
Interestingly, the coordinates of the entire compositional table can be divided into two groups according to the dimensionality of the independence and interaction tables, respectively. This becomes the main advantage also when using pivot coordinates for robust PCA since it allows for a comparison of the results from the whole table and its decomposed parts. 

\indent Generally, there are three types of pivot coordinates corresponding to the row, column and ``odds ratio'' partitioning of the compositional table \cite{facevicova16}. 
The first two types together form a coordinate representation of the independence table, the third one is used for the interaction table. Altogether, they form a coordinate representation of the original compositional table. In case of row and column types of coordinates, the entire first row or column, respectively, is taken as the pivot element and separated from the rest. In the next step, this pivot is not considered anymore and the following row or column is taken as the new pivot element, and so on, until the following $I+J-2$ coordinates are obtained, 

\begin{equation}
\label{eq:ind1}
 z^r_i=\sqrt{\frac{(I-i)J}{1+I-i}}\ln\frac{g(\boldsymbol{x}_{i\bullet})}{[g(\boldsymbol{x}_{i+1\bullet}),...,g(\boldsymbol{x}_{I\bullet})]^{1/(I-i)}},\quad i=1, ..., I-1,
\end{equation}

\begin{equation}
\label{eq:ind2} z^c_j=\sqrt{\frac{I(J-j)}{1+J-j}}\ln\frac{g(\boldsymbol{x}_{\bullet j})}{[g(\boldsymbol{x}_{\bullet j+1}),...,g(\boldsymbol{x}_{\bullet J})]^{1/(J-j)}},\quad \quad j=1, ..., J-1,
\end{equation}
where $g(\boldsymbol{x}_{i\bullet})$ and $g(\boldsymbol{x}_{\bullet j})$ stand for the geometric mean of the $i$-th row and $j$-th column, respectively. 

The process of obtaining the remaining $(I-1)(J-1)$ coordinates is based on a division of the original compositional table into four blocks, say upper left A, upper right B, lower left C and lower right D, where A contains always just one (pivot) cell indexed by $rs$. The odds ratio interpretation should be now easily seen from the following formula, where the elements of blocks A and D are in the numerator, and the elements of blocks B a C in the denominator of the logratio;
\begin{equation}
\label{eq:int1}
 z^{OR}_{rs}=\sqrt{\frac{1}{(I-r)(J-s)(I-r+1)(J-s+1)}}\ln\prod_{i=r+1}^I \prod_{j=s+1}^J \frac{x_{ij}x_{rs}}{x_{is}x_{rj}}.
\end{equation}
To obtain all coordinates of the odds ratio type in a proper order corresponding to the $z^r$ and $z^c$ coordinates, the position of the pivot cell is moving firstly by rows with fixed first column, $r=1, \ldots, I-1$, then by columns with fixed last row, $s=1, \ldots, J-1$, and afterward the row position is always leveled back down by one and the column position moves again from $1$ to $J-1$ for the given row until all sizes of the $r \times s$ table are covered.

For the sake of completeness, permutations of the entire rows or columns following the same principle as stated in Section~\ref{sec:2} could be performed. Hereby for all combinations of rows and columns, different coordinate systems consisting of $z_i^{r(k)}$, $z_j^{c(l)}$ and $z_{rs}^{OR(kl)}$, where $(kl), k=1, \ldots, I, l=1, \ldots, J,$ defines row and column permuted to the pivot position within the whole table, would be gained \cite{facevicova16}. 
Following (\ref{clrscaled}), also the first coordinates of the three types from each system can then be expressed as proportional (up to a constant) to respective clr coefficients, 
\begin{equation}\mathrm{clr}(\boldsymbol{x}_{ind})_{kl}=\sqrt{\frac{I-1}{IJ}}z_1^{r(k)}+\sqrt{\frac{J-1}{IJ}}z_1^{c(l)},
\end{equation}
\begin{equation}
\mathrm{clr}(\boldsymbol{x}_{int})_{kl}=\sqrt{\frac{(I-1)(J-1)}{IJ}}z_{11}^{OR(kl)},
\end{equation}
which is an important fact for the interpretation of the analysis. The resulting clr coefficients, computed originally from the elements of the independence and interaction tables, 
\begin{equation} \mathrm{clr}(\boldsymbol{x}_{ind})_{ij}=\ln \frac{x_{ij}^{ind}}{g(\boldsymbol{x}_{\bullet \bullet}^{ind})}, \quad
\mathrm{clr}(\boldsymbol{x}_{int})_{ij}=\ln \frac{x_{ij}^{int}}{g(\boldsymbol{x}_{\bullet \bullet}^{int})},
\end{equation}
can thus be expressed also in terms of cells of the input compositional table as
\begin{equation} 
\label{clrgmean}
\mathrm{clr}(\boldsymbol{x}_{ind})_{ij}=\ln \frac{g(\boldsymbol{x}_{i \bullet})g(\boldsymbol{x}_{\bullet j})}{g(\boldsymbol{x}_{\bullet \bullet})^2}, \quad
\mathrm{clr}(\boldsymbol{x}_{int})_{ij}=\ln \frac{x_{ij}g(\boldsymbol{x}_{\bullet \bullet})}{g(\boldsymbol{x}_{i \bullet})g(\boldsymbol{x}_{\bullet j})},
\end{equation}
\noindent where $g(\boldsymbol{x}_{i\bullet})$, $g(\boldsymbol{x}_{\bullet j})$ and $g(\boldsymbol{x}_{\bullet \bullet})$ stand for the geometric mean of the $i$-th row, the $j$-th column and the whole compositional table (and its independent and interactive counterparts), respectively. As a consequence, each $\mathrm{clr}(\boldsymbol{x}_{ind})_{ij}$ expresses a~dominance of a~given combination of factor values in case of independence. This dominance is then either amplified or weakened according to the interaction table which depends on whether the interaction is shifted in a positive or a negative direction. The interaction table refers also to sources of departures from independence, nevertheless, the information obtained only from $\mathrm{clr}(\boldsymbol{x}_{int})_{ij}$ does not provide a complete picture about the dominance of the respective cell to all other averaged cells. 

\indent Furthermore, note that each coordinate $\mathrm{clr}(\boldsymbol{x}_{ind})_{ij}$ is formed by the sum of clr coefficients of the respective row and column marginals, $1/J\sum_j{\mathrm{clr}(\boldsymbol{x}_{ind})_{ij}} \\ =\ln g_{i\bullet}/g_{\bullet\bullet}$ and $1/I\sum_i{\mathrm{clr}(\boldsymbol{x}_{ind})_{ij}}=\ln g_{\bullet j}/g_{\bullet\bullet}$, which amount to zero. Thus, there are only $I+J-2$ linearly independent clr coefficients, reflecting the dimensionality of the sample space of independence tables again. A similar feature holds also for clr coefficients of interaction tables that sum up to zero across each row or column. Consequently, in the case of an interaction table, the number of linearly independent clr coefficients reduces to $(I-1)(J-1)$. Since this dependency makes it impossible to use the clr coefficients for the robust PCA of independence and interaction tables,
the strategy to perform robust PCA for compositional tables is the same as in case of vector compositional data: PCA loadings and scores are computed in ilr coordinates and then back-transformed using relation (\ref{ilrclr}) to the clr space, where the loadings can be interpreted in terms of dominance of single cells. Here, clr coefficients of basis vectors for rows $\boldsymbol{e}^r$, columns $\boldsymbol{e}^c$ and interactions $\boldsymbol{e}^{OR}$, forming the columns of the matrix $\boldsymbol{V}$, are defined as follows,

\begin{equation}
\label{row}
\mathrm{clr}(\boldsymbol{e}^r)=
\begin{cases} \sqrt{\frac{I-i}{(I-i+1)J}} & \text{for the elements in pivot row } i, \\
-\sqrt{\frac{1}{(I-i+1)J(I-i)}} & \text{for the elements in rows } i+1, ..., I, \\
0 & \text{otherwise},
\end{cases}
\end{equation}

\begin{equation} 
\label{column}
\mathrm{clr}(\boldsymbol{e}^c)=
\begin{cases} \sqrt{\frac{J-j}{(J-j+1)I}} & \text{for the elements in pivot column } j, \\
-\sqrt{\frac{1}{(I-i+1)J(I-i)}} & \text{for the elements in columns } j+1, ..., J, \\
0 & \text{otherwise},
\end{cases}
\end{equation}

\noindent and

\begin{equation} 
\label{interact}
\mathrm{clr}(\boldsymbol{e}^{OR})=
\begin{cases} \sqrt{\frac{1}{rs(r-1)(s-1)}} & \text{for the elements on positions } i=r+1, ..., I, \\
 & j=s+1, ..., J \\
\sqrt{\frac{(r-1)(s-1)}{rs}} & \text{for the pivot elements } rs \\
-\sqrt{\frac{r-1}{rs(s-1)}} & \text{for the elements in pivot row } r, \\
 & j=s+1, ..., J, \\
-\sqrt{\frac{s-1}{rs(r-1)}} & \text{for the elements in pivot column } s, \\
 & i=r+1, ..., I, \\
0 & \text{otherwise},
\end{cases}
\end{equation}
reinterpreting the expressions from \cite{facevicova16}. 
As a result of (\ref{ilrclr}), row-wise clr coefficients of the whole table are obtained for the $IJ-1$ columns of the matrix $\boldsymbol{V}$. Alternatively, if the matrix $\boldsymbol{V}$ has just $I+J-2$ columns formed by clr coefficients of basis vectors corresponding to the ilr representation of the independence table (\ref{eq:ind1}), (\ref{eq:ind2}), its respective clr coefficients are derived (and similarly for the interaction table with its coordinates (\ref{eq:int1})). Finally, the transformed loadings and scores can be used to construct a biplot in order to reveal the multivariate structure of the sample of compositional tables and relations between both factors.

\section{Unemployment data analysis}
\label{sec:5}

In the following, the methodological results are applied to two real-world data sets in order to illustrate the main features and possible limitations of the approach.
In the first case, data from OECD Statistics about more than 150 million unemployed people from 42 different countries in 2010 \cite{oecd17a} 
are analyzed using the statistical software environment R \cite{r18}. 

The data set contains the numbers of unemployed people together with their gender and age category for the following countries: Australia, Austria, Belgium, Canada, Chile, Czech Republic, Denmark, Estonia, Finland, France, Germany, Greece, Hungary, Iceland, Ireland, Israel, Italy, Japan, Korea, Mexico, Netherlands, New Zealand, Norway, Poland, Portugal, Slovakia, Slovenia, Spain, Sweden, Switzerland, Turkey, United Kingdom, United States, Colombia, Costa Rica, Latvia, Lithuania, China, India, Indonesia, Russian Federation and South Africa. An example of (transposed) raw data from the first four countries is shown in Table~\ref{tab:1}. The numbers in the tables are basically counts of unemployed people according to two factors. 
As the population size varies among the countries, the interest
here is not in the absolute values of the counts in the single countries, but rather the relative structure of unemployment. Particularly, ratios of men and women and ratios among age groups 15-24, 25-39, 40-54 and 55+, as well as proportionality among countries will be analyzed. 
Since outliers can be anticipated due to completely different economics, education levels, gender balance and also traditions of the listed countries, the analysis will be carried out in a robust manner. 

\begin{table}
{\begin{tabular*}{\textwidth}{l@{\extracolsep{\fill}}llllllllll}
                           & \multicolumn{2}{l}{Australia} & \multicolumn{2}{l}{Austria} & \multicolumn{2}{l}{Belgium} & \multicolumn{2}{l}{Canada}  \\ \hline
\multicolumn{1}{l|}{Age}   & Men          & Women          & Men         & Women         & Men         & Women         & Men         & Women                \\ \hline
\multicolumn{1}{l|}{15-24} & 129          & 111            & 29          & 25            & 53          & 43            & 250         & 178                    \\
\multicolumn{1}{l|}{25-39} & 90           & 85             & 40          & 35            & 86          & 83            & 241         & 192                  \\
\multicolumn{1}{l|}{40-54} & 66           & 68             & 36          & 27            & 65          & 52            & 242         & 188                     \\
\multicolumn{1}{l|}{55+}   & 37           & 19             & 7           & 3             & 13          & 11            & 121         & 75        
                     \\
\hline
\end{tabular*}}
\caption{Unemployed people in thousands partitioned according to their gender and age groups \cite{oecd17a}.} 
\label{tab:1}
\end{table}

\indent All compositional tables in this example have two rows and four columns, i.e., gender is the row factor and age structure is the column factor. The sample space of tables thus has dimension $7$ out of which independence tables account for a dimension of $4$ with pivot coordinates $z_1^r$, $z_1^c$, $z_2^c$ and $z_3^c$, while the remaining coordinates $z_{11}^{OR}$, $z_{12}^{OR}$ and $z_{13}^{OR}$ correspond to the interaction tables with a dimension of $3$. 

\indent To point out the differences between the classical and robust PCA, both are performed and compared through the resulting covariance compositional biplots. Recall that classical PCA can directly be applied on clr coefficients. Nevertheless, since for this data set robust PCA may be more relevant because of potential outlying tables, ilr coordinates are used in both cases, and the results are transformed to clr for the biplot construction. This can only result in a different rotation of the classical biplot, however, it obviously does not alter the results.   

\begin{figure}
\centering
\includegraphics[width=0.88\linewidth]{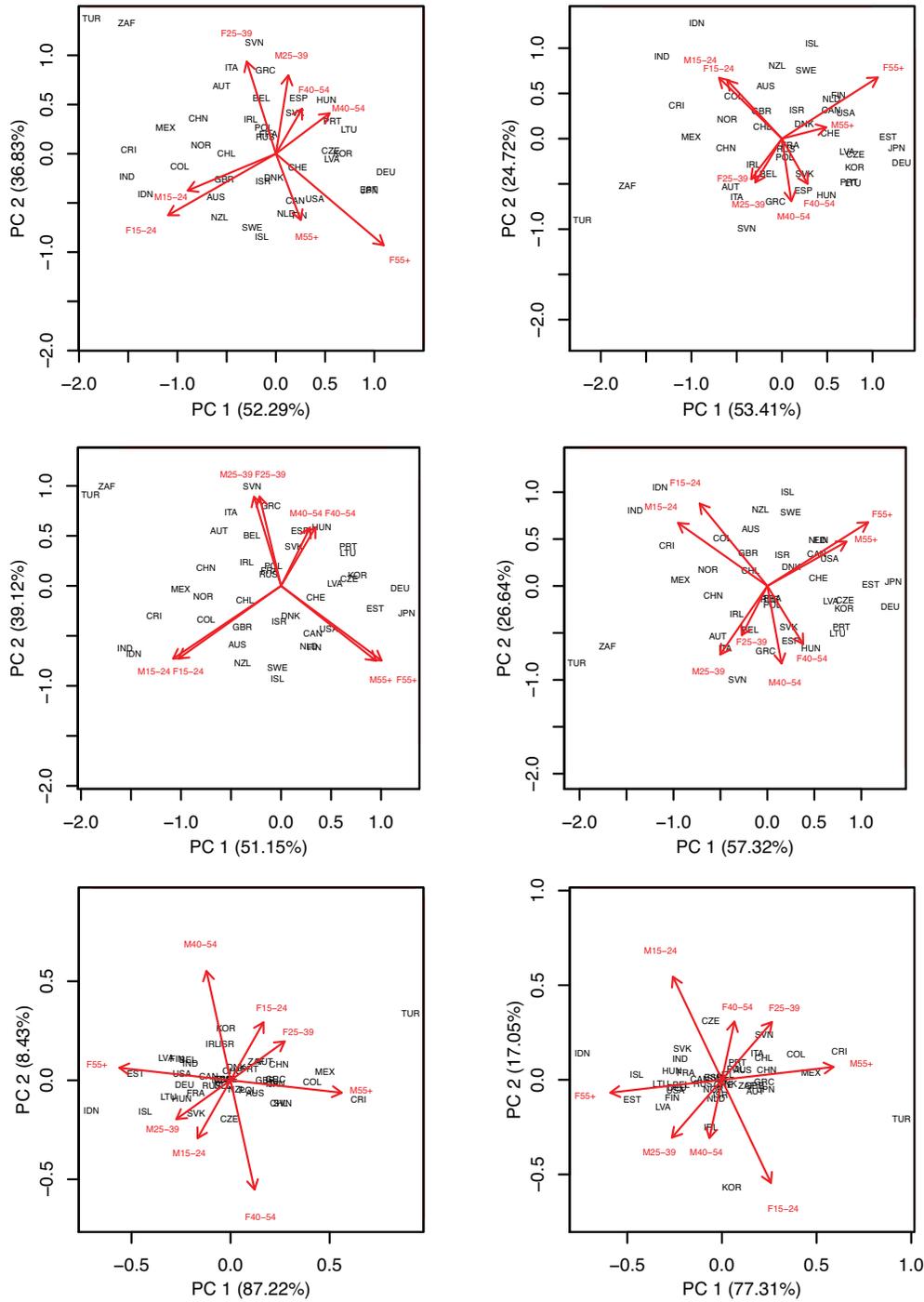}
\vspace{-13mm}
\caption{Robust (left column) and classical (right column) covariance biplots of the Unemployment compositional (upper row), independence (middle row) and interaction tables (lower row).}
\label{fig:1}
\end{figure}

\indent In order to perform PCA in ilr coordinates, the standard function \texttt{princomp} in R can be used, where the parameter \texttt{covmat} is set to \texttt{covMcd} (MCD estimator of covariance) in case of robust PCA. Thereafter, loadings need to be transformed to clr coefficients as described in Section~\ref{sec:2} using the matrix $\boldsymbol{V}$ with columns defined by (\ref{row}) - (\ref{interact}) for the entire compositional table, and by (\ref{row}) and (\ref{column}), or by 
(\ref{interact}) for its independent and interactive part, respectively. The resulting classical biplots are depicted on the right-hand side of Figure~\ref{fig:1}, while the robust PCA output is on the left.

Assessing Figure~\ref{fig:1}, it can be noticed that in all three cases, robust PCA performs better in terms of explained variability by the first two principal components. 
As mentioned above, some outliers might be present in the data, and even from the classical biplot of the whole compositional tables (upper right corner of Figure~\ref{fig:1}), at least two outlying tables (Turkey-TUR and South Africa-ZAF) could be expected, and thus the robust approach should provide more meaningful results. 
However, an outlier detection is performed additionally in order to confirm these expectations. In the R package robCompositions \cite{templ11}, 
there is a function \texttt{outCoDa} defined for this purpose, based on robust Mahalanobis distances computed from ilr transformed data \cite{filzmoser08}. 
Using the pivot coordinates and applying the $0.975$ quantile of the chi-squared distribution as the common cut-off value, 15 out of all 42 countries are identified as outlying observations, clearly supporting the choice of robust analysis. Note that similarly, 10 observations from the set of independence tables and 6 from the interaction tables were detected as potential outliers.  

\indent Additionally, from the same part of the figure, it is easy to identify from the direction of the arrows which countries tend to have relatively higher unemployment among younger people and which ones have a rather higher rate in the opposite situation. Although no clear compact clusters are visible, it is obvious that most European countries together with the USA and Canada tend to have most likely problems with employing older people, say 40+, while for Central and South America together with China, India, and Indonesia the unemployment depending on age structure has rather opposite tendencies. Moreover, also some gender differences can be observed, except for the youngest generation. The structure in the classical biplot (upper right plot) is similar but driven by the identified outlying observations.

The left plot in the middle part of Figure~\ref{fig:1} shows the ``ideal'' situation in case the relationships between gender and age factors would be filtered out. While the positions of the countries are not apparently changed compared to the previously discussed covariance biplot (upper right corner), the general relationships between the factors are remarkably illustrative. In case of independence, nearly gender equity would be achieved, while on the contrary, relationships among the age levels would be disproportionally weaker. Also, a bigger difference between results provided by robust and classical PCA is present here. One can easily understand how the classical approach does not handle outliers and how those can affect the output; the biplot on the right side is quite far away from picturing the same ideal situation.

As demonstrated, the independence table captures the hypothetical balanced state with each clr interpreted in terms of dominance of a given combination of factors in case of independence. However, this dominance is then either amplified, or weakened according to the interaction table; in terms of clr coefficients, it depends on whether the logratio dominance is shifted in a positive, or in a negative direction. Note that information obtained from the interaction table only does not provide a complete picture about the dominance of single cells in the table. For example, in the lower left biplot (robust biplot of interaction tables), Costa Rica (CRI) is placed towards the loading ``male 55+'', but this does not necessarily lead to a conclusion that unemployment in this group is higher in general in this country; it simply marks the cell whose dominance causes imbalance for Costa Rica, although the actual proportion of unemployment for this age group can be lower than its average dominance. Therefore, the conclusion about the higher dominance of unemployed men in the oldest group than expected in the hypothetical case of independence can be stated only after looking at the biplot of the independence tables. For the compositional tables with dimension $2 \times J$ (or alternatively 
$I\times 2$), this feature is nicely illustrated by the depicted loadings themselves, placed along a line corresponding to increasing dominance of one factor value at the expense of the latter value of the same factor. The opposite relation between the respective clr coefficients is clearly visible from both the biplots and the form of $\mathrm{clr}(\boldsymbol{x}_{int})_{ij}$ in (\ref{clrgmean}): as it was already discussed in Section~\ref{sec:4}, clr coefficients of the interaction table sum up to zero across each row and column which results in the identity $\mathrm{clr}(\boldsymbol{x}_{int})_{1j}=-\mathrm{clr}(\boldsymbol{x}_{int})_{2j}$, holding for each $j$ when $I=2$ (and similarly for $J=2$). 
While in case of higher data dimension the property is no longer visible in the graphs, in this example it can be seen that the two possible values of the gender factor lead to precisely contradictory loadings for any chosen value of the age factor. Thus they might only carry the information about the origin of the dominance shift, but no longer about the direction of the shift for which the difference from the independence table has to be consulted.

\section{Education data analysis}
\label{sec:6}

It was illustrated in the previous example how outliers can affect results of classical PCA. Especially the gender equity achieved in the robust biplot of the independence tables would not be present in the classical one. Hence, in this second example, only robust analysis outputs are discussed.
The data set contains information about more than 7 million female and nearly 6 million male students, divided according to 8 different fields of study, being Education, Humanities and arts, Social sciences, business and law, Science, mathematics and computing, Engineering, manufacturing and construction, Agriculture and veterinary, Health and welfare, and Services \cite{oecd17b}. 
The information about the achieved degree (bachelor, master, and doctoral, respectively) is recorded as well for about 30 different countries.

\begin{figure}
\centering
\includegraphics[width=0.88\linewidth]{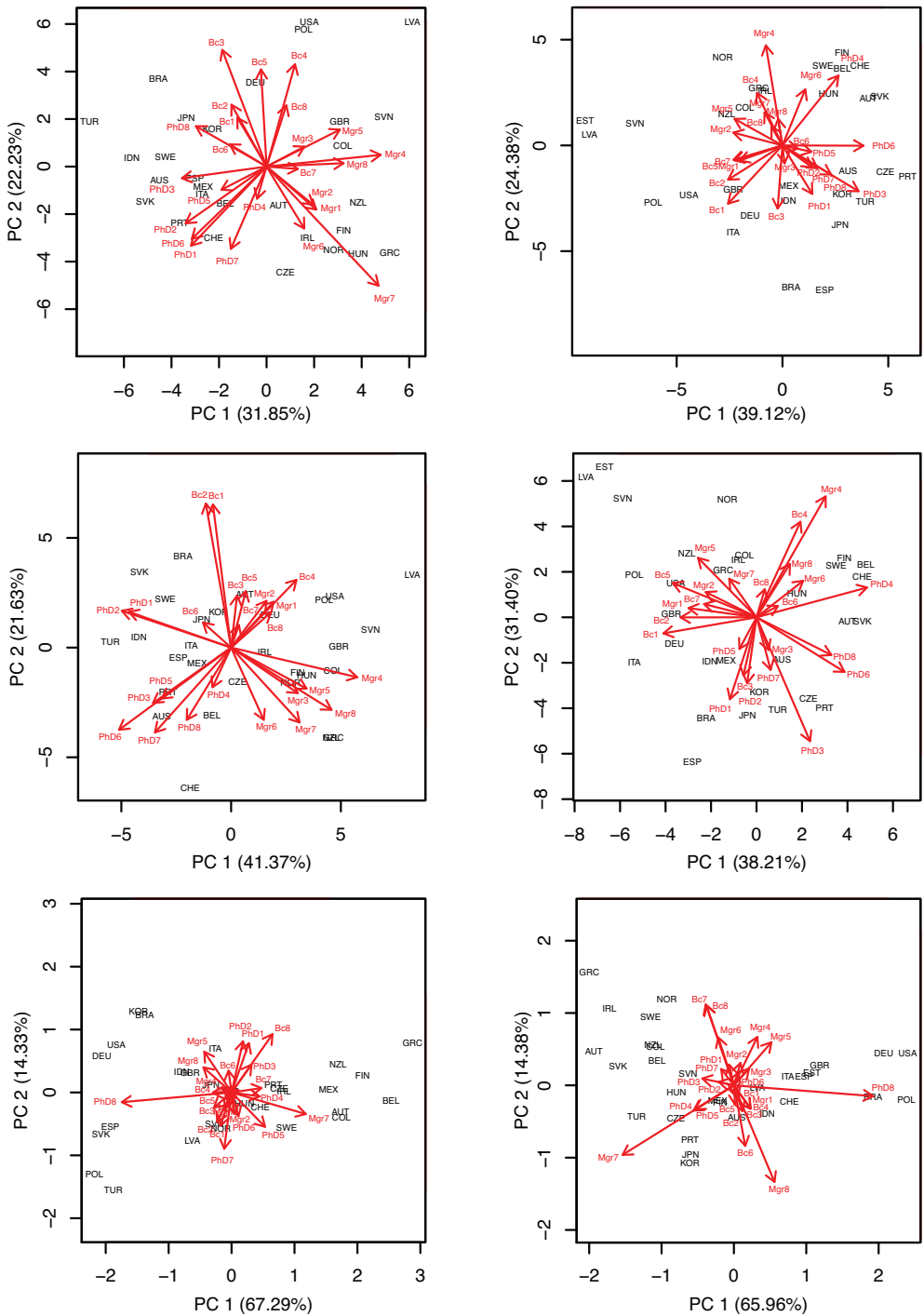}
\vspace{-13mm}
\caption{Robust covariance biplots of the Education compositional (upper row), independence (middle row) and interaction tables (lower row) for men (left column) and women (right column), respectively. Study fields are marked as follows: $1 =$ Education, \mbox{$2 =$ Humanities} and arts, $3 =$ Social sciences, business and law, $4 =$ Science, mathematics and computing, $5 =$ Engineering, manufacturing and construction, $6 =$ Agriculture and veterinary, $7 =$ Health and Welfare, and $8 =$ Services.}
\label{fig:2}
\end{figure}

Compositional tables are analyzed for both genders separately in order to allow for a comparison of possible differences between them later on. Biplots as graphical PCA outcomes of the whole compositional table as well as independence and interaction tables are collected in Figure~\ref{fig:2}. Due to a larger dimensionality of compositional tables than in previous case ($3 \times 8$), the biplots contain three times more variables and an objective interpretation becomes more difficult. An additional aspect is that since it is necessary to go many dimensions down to achieve the PCA projection using the first two principal components, it is expected to obtain more approximative picture of the multivariate data structure in the biplot. However, for data of similar or even bigger size, the proposed methods still offer an extremely useful rank-two approximation capturing the relationships between both factors. The performance of robust PCA is still good enough (at least $54.08 \%$) for this concrete case in terms of cumulative variability explained by the first two principal components.

Despite the previous interpretational doubts, it can be seen that the effect of the chosen final degree is possibly stronger than the effect of the study field since the loadings tend to create a quite clear division of bachelors, masters, and doctors for most of the biplots. 
This property is more obvious for men while for women the difference between bachelors and masters is partially wiped out, maybe also due to a less compact data structure. Finally, employing the \texttt{outCoDa} function \cite{templ11,filzmoser08} again, some outliers for the data set of the whole compositional tables are detected: Austria, Norway and Spain for men, and United Kingdom, Turkey and United States for women.

From the second row of the figures, some overall idea about the hypothetical state of independence between degree and study field factors might be obtained.  
A stronger effect of the chosen degree and a weaker effect of the study field would still be apparent for men, and one new feature could be observed: there would be a strong relation between Education and Humanities and arts study fields for each degree. For women, e.g. a~similarity of educational systems in Sweden, Finland, Belgium and Switzerland is reflected. Also, in the case of independence, higher occurrence of outliers would be present for both men and women.

When looking at the biplot for the interaction tables for women (lower right figure), two of the mentioned countries are shifted away from the fields of study that would dominate if independence was achieved, being especially Agriculture and veterinary, Science, mathematics and computing, and Services. These countries are Sweden and Belgium, while Finland would correspond approximately to the independence between the factors, and Switzerland actually accounts for even stronger dominance of those fields (particularly master and doctoral studies in Services). For both men and women, it could be stated that the actual relationships between the factors are quite distant from the relative dominance given in the independence state. Stronger patterns concerning both factors are generally observed for men.

\section{Conclusions}
\label{sec:7}

Vector compositional data, or their generalization to compositional tables, are characterized by their relative scale and scale invariance properties captured by the Aitchison geometry. Accordingly, a representation of the compositional tables isometrically in clr coefficients or ilr coordinates is essential for the further statistical analysis using popular multivariate methods. Compositional tables can be decomposed onto their independence and interaction parts; a statistical analysis of both is recommended to get insight into the ideal situation when relationships between factors are filtered away, as well as into interactions between factors forming the compositional table. As most real-world datasets contain outlying observations, robust methods requiring an orthonormal coordinate representation have been considered. To reduce the dimension of data at hand, a robust PCA using the MCD estimator can be applied to pivot coordinates of compositional, independence and interaction tables (but also any other ilr coordinates which respect the decomposition of compositional tables could be used). The necessity of respecting dimensionality of independence and interaction tables forms the main difference to (vector) compositional data where such feature does not occur.
In case of pivot coordinates, it is also possible to find a simple relationship between them and the respective clr coefficients as well as to enhance interpretation of the latter variables. Therefore, PCA loadings obtained in pivot coordinates are transformed back to clr coefficients where they are used for the construction of biplots and their meaningful analysis. In case of $(2 \times J)$ table dimensions, an additional feature can be observed in the graphical output of interaction tables, which can be traced back to the interpretation of the clr coefficients. As a next step, three factors of a compositional cube $(I \times J \times K)$ could be analyzed by robust PCA and appropriate 3D graphical tools employed. Consequently, the educational datasets for men and women could be merged into one and approached accordingly. The case of compositional cubes and also its further extension to generally $N$ factors belong to primary research interests of the authors.

\section*{Funding}

This work was supported by COST Action CRoNoS under Grant IC1408; The Czech Science Foundation under Grant 18-05432S; and the grant IGA\_PrF\_2018\_024 Mathematical Models of the Internal Grant Agency of the Palack\'y University in Olomouc.

\end{document}